\documentclass[ aip,pof,amsmath,amssymb,reprint,onecolumn]{revtex4-1}

\usepackage{graphicx}
\usepackage{dcolumn}
\usepackage{bm}
\usepackage{geometry}
\begin{document}


\title[Damping of liquid sloshing by foams]{Damping of liquid sloshing by foams}

\author{A. Sauret}\email{alban.sauret@saint-gobain.com}
\affiliation{Department of Mechanical and Aerospace Engineering, Princeton University, Princeton, NJ 08544, USA}
\affiliation{Surface du Verre et Interfaces, UMR 125 CNRS/Saint-Gobain, 93303 Aubervilliers, France.}

\author{F. Boulogne}
\altaffiliation{A. Sauret and F. Boulogne contributed equally to this work.}
\affiliation{Department of Mechanical and Aerospace Engineering, Princeton University, Princeton, NJ 08544, USA}

\author{J. Cappello}
\affiliation{Department of Mechanical and Aerospace Engineering, Princeton University, Princeton, NJ 08544, USA}
\affiliation{Ecole Normale Sup\'erieure de Cachan, 94235 Cachan, France}

\author{E. Dressaire}
\affiliation{Department of Mechanical and Aerospace Engineering, Princeton University, Princeton, NJ 08544, USA}
\affiliation{Department of Mechanical and Aerospace Engineering, New York University Polytechnic School of Engineering, Brooklyn, NY 11201, USA}

\author{H. A. Stone} \email{hastone@princeton.edu}
\affiliation{Department of Mechanical and Aerospace Engineering, Princeton University, Princeton, NJ 08544, USA}

\date{\today}

\begin{abstract}
When a container is set in motion, the free surface of the liquid starts to oscillate or slosh. Such effects can be observed when a glass of water is handled carelessly and the fluid sloshes or even spills over the rims of the container. However, beer does not slosh as readily as water, which suggests that foam could be used to damp sloshing. In this work, we study experimentally the effect on sloshing of a liquid foam placed on top of a liquid bath. We generate a monodisperse two-dimensional liquid foam in a rectangular container and track the motion of the foam. The influence of the foam on the sloshing dynamics is experimentally characterized: only a few layers of bubbles are sufficient to significantly damp the oscillations. We rationalize our experimental findings with a model that describes the foam contribution to the damping coefficient through viscous dissipation on the walls of the container. Then we extend our study to confined three-dimensional liquid foam and observe that the behavior of 2D and confined 3D systems are very similar. Thus we conclude that only the bubbles close to the walls have a significant impact on the dissipation of energy. The possibility to damp liquid sloshing using foam is promising in numerous industrial applications such as the transport of liquefied gas in tankers or for propellants in rocket engines.
\end{abstract}
%

\maketitle


\section{Introduction}

Sloshing is the motion of the free surface of a liquid induced by variations of the velocity of the container.
The amplitude of the generated waves depends on the frequency and the amplitude of motion of the container.\cite{McIver1989,Ibrahim,Molin2001,Weidman2012,Hopfinger2012,Timokha2012,Turner2013,Bouscasse2014a,Reclari2014} When the frequency of the motion corresponds to the resonance frequency of the surface wave, the amplitude increases significantly.  If the amplitude of sloshing becomes large enough, spilling, splashing, and/or drop formation are possible.\cite{Mayer2012} Such phenomenon can lead to challenging technical constraints in various applications including the transportation of oil and liquefied gas in tankers. Indeed, the sloshing dynamics leads to considerable pressure forces on the walls of a container and can be a source of destabilization or failure of a container. Therefore, the characterization of the amplitude of the waves and the exploration of mechanisms capable of damping sloshing is of critical importance. To damp sloshing, various methods have been considered including the use of elastic membranes attached to the top of a container, \cite{Bauer1995,Bauer2007,Kidambi2009} or baffles placed in a container.\cite{Maleki2008,Liu2009} The use of containers with elastic walls was also considered.\cite{Bauer1968} However, these methods can be difficult to implement in large containers or require the design of new containers.

In the present paper, we study the damping effects of a liquid foam placed on top of the liquid. The present method is versatile as it can be used in any container. Foam is easy to produce and cheap as it requires a small quantity of foaming solution to generate a large volume of foam. In addition, foams are already used in various industrial applications and processes such as oil recovery, soil remediation, fire containment or blast wave mitigation.\cite{Prudhomme1995,Stevenson2012}

A foam, as used in the present study, is an assembly of densely packed bubbles produced upon aeration of a surfactant solution. The flow of foam in two-dimensional channels has been studied extensively both experimentally\cite{Cantat2004,Dollet2007,Dollet2010,BenSalem2013} and numerically. \cite{Cox2004,Cox2005,Boulogne2011}
Common foams exhibit viscous, elastic and/or plastic responses depending on the external forcing. \cite{Bronfort2014,Marmottant2007,benito} For small deformations, foams behave as visco-elastic solids.\cite{Cohen-Addad2013} In this regime, the effective viscosity of a foam is much larger than the viscosity of the foaming solution.\cite{mousses,rheo} Therefore, foam is a good candidate to damp sloshing and thus to decrease the pressure forces exerted on the walls of a container.

Since the seminal work of Bretherton on the motion of a bubble in a capillary,\cite{Bretherton1961} the viscous dissipation of a flowing foam has been considered.\cite{Denkov2005,Saugey2006,Raufaste2009,Costa2013,Cantat2013,LeMerrer2014}
It has been shown that the friction coefficient depends on the capillary number (ratio of viscous forces to surface tension forces) but varies with the nature of the foam. For three-dimensional foams, two sources of dissipation need to be taken into account. The first contribution is the viscous friction induced by the displacement of the Plateau borders associated with the displacement of the bubbles in contact with the walls of a container and relative to one another. The second contribution depends on the interfacial rheology of the foam and is induced by the presence of surfactant molecules at air/liquid interfaces.\cite{Denkov2005,Cantat2013} To characterize the relative magnitude of the different contributions, a distinction should be made between low and high surface moduli. At high surface modulus, interfaces behave as incompressible surfaces.\cite{Langevin2014} In the present study, we rely on a foam with low surface modulus where the interfaces are elastic.

In this paper, we study the damping effect of a foam placed on top of a liquid bath in an oscillating rectangular container. In section \ref{sec:experiment} we begin by presenting the experimental setup, the physical properties of the system and the two methods of excitation considered. Then, in section \ref{sec:newtonian} we characterize the sloshing of a single Newtonian fluid. We recall analytical results previously derived and validate the experimental approach. We also introduce the damping coefficient $1/\tau$ and provide a scaling law to rationalize the experimental results. Section \ref{sec:foam} is devoted to the sloshing in presence of a layer of foam of varying thickness. We extend our observations to three-dimensional foams and present a phenomenological model that predicts the damping of the wave as a function of the foam thickness. We summarize our conclusions and discuss avenues for future research in section \ref{sec:conclusion}.

\section{Experimental methods}\label{sec:experiment}

\subsection{Experimental setup}

\begin{figure}
    \centering
    \includegraphics{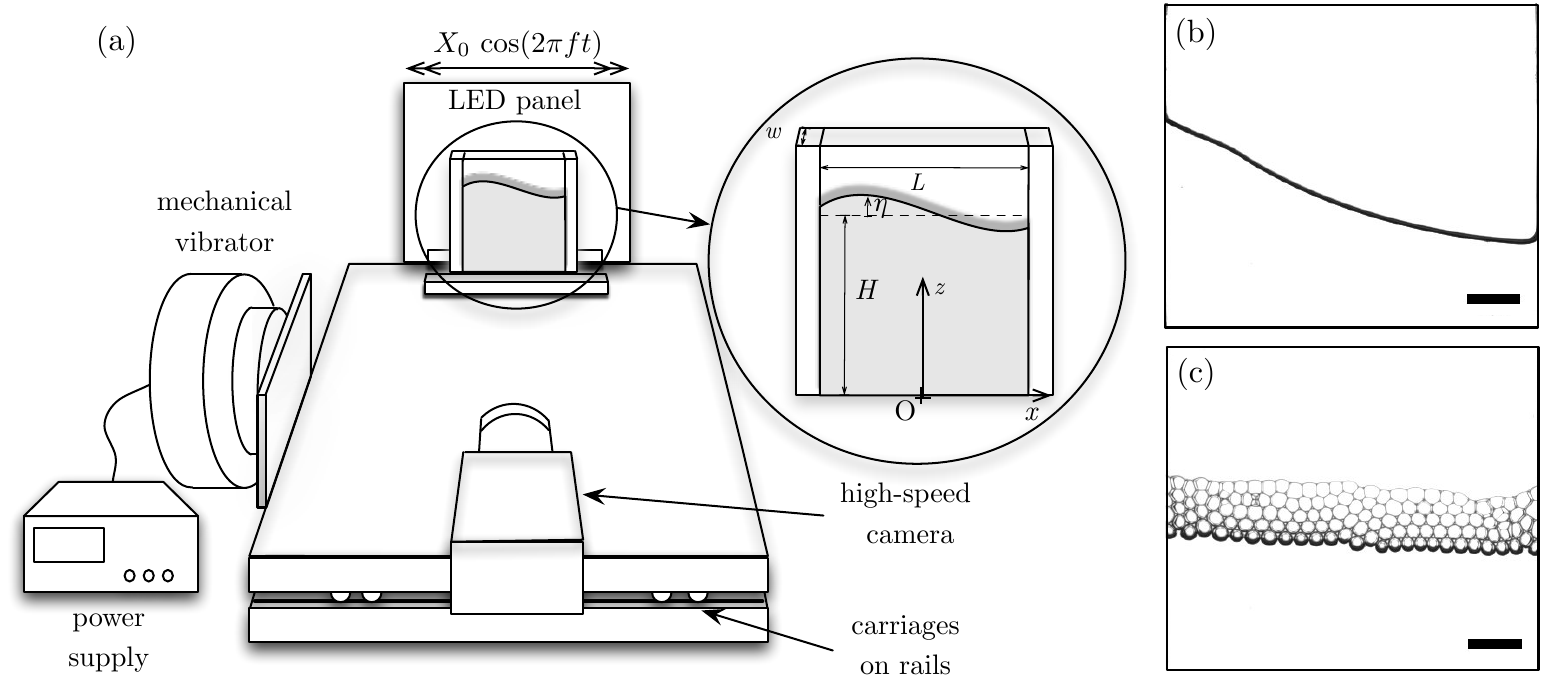}
    \caption{(a) Schematic of the experimental setup. Example of visualization for (b) a Newtonian fluid and for (c) a Newtonian fluid with a $2$D liquid foam on the top. The images are recorded at the maximum amplitude of the oscillation. Scale bars are $1$ cm.}\label{fig:setup}
\end{figure}

To perform visualization of the different interfaces (liquid/air, liquid/foam, foam/air) and track the motion of the foam, we use a rectangular cell in which the displacement of the fluid remains mainly two-dimensional. The vertical walls of the container are made of borosilicate glass (purchased from Ted Pella and McMaster-Carr) and a rigid rubber sheet (McMaster-Carr) is used to close the bottom of the cell. This configuration allows the introduction of a needle at the center of the container to generate the monodisperse foam. The cell has a length of $L=70\,\rm{mm}$, a height of $92\,\rm{mm}$ and a varying width $w\in[3;\,19]\,\rm{mm}$.
The experimental apparatus, represented in figure \ref{fig:setup}(a), consists of a container, a LED panel and a high-speed camera mounted on a moving stage. The moving stage is on rails and set in motion by a mechanical vibrator (LDS 319024-3) controlled by a power supply (Stanford research system Model DS345) and an amplifier (LDS PA100E). The lighting is provided by the LED panel (Phlox, 10 cm $\times$ 10 cm) placed behind the cell. The oscillations of the liquid surface are recorded using a high-speed camera  (Phantom V9.1) operating at typically 100 frames per second. Examples of the visualizations are shown without and with foam, respectively, in figures \ref{fig:setup}(b) and \ref{fig:setup}(c).

We first study the sloshing dynamics of a single Newtonian fluid. We use different fluids to vary the viscosity, the density and the surface tension of the liquid. We compare the behavior of water, mixtures of water and glycerol (50\% v/v, 60\% v/v and 80\% v/v of glycerol), dodecane, decanol and a foaming solution (described in detail in the next paragraph). We are therefore able to study how the physical parameters of the system influence the sloshing dynamics of a single Newtonian fluid. The viscosity and the surface tension of the Newtonian fluids are measured, respectively, with an Anton Paar rheometer (MCR 301, CC50 geometry) and the pendant drop method at $20^{\rm o}{\rm C}$ and compared with tabulated values.\cite{Sheely1932,Lide2008,Soller2008} The physical properties of the different fluids used in this study are summarized in Table \ref{tab:fluides}.

\begin{table}
    \begin{tabular}{|c|c|c|c|}
        Fluid & kinematic viscosity $\nu$ & surface tension $\gamma$  & density $\rho$ \\
              & ($\rm{\times 10^{-6}  \, {m}^{2}\,{s^{-1}}})$  & $\rm{(\times 10^{-3} \,\, N\,m^{-1})}$ & $\rm{(\times 10^{3}   \,\,kg\,m^{-3})}$\\
        \hline
        water &  $1.01 $ & $72$ & $1$ \\
        water-glycerol (50 vol\%) & $5$ & $67$ & 1.13\\
        water-glycerol (60 vol\%) & $9$ & $67 $ & 1.15\\
        water-glycerol (70 vol\%) & $19$ & $66 $ & 1.18\\
        water-glycerol (80 vol\%) & $46$ & $66 $  & 1.21\\
        foam solution & $1.4$ & $25$ &  $1.02 $ \\
        dodecane &  $1.8$ & $25$ & $0.75$ \\
        decanol & $13.1$  & $28$  & $0.83$  \\
    \end{tabular}
    \caption{Physical properties of the fluids measured at $20^{\rm o}\,{\rm C}$.}\label{tab:fluides}
\end{table}

To perform sloshing experiments in the presence of foam, we partially fill the cell with a foaming solution well above the critical micelle concentration (CMC). This solution is made of 90\% v/v water, 5\% v/v glycerol and 5\% v/v of commercial surfactant (Dawn dish-washing liquid, Procter \& Gamble). The surfactant molecules lead to a stable foam with low surface modulus.\cite{Denkov2009} Glycerol is used to slow down the aging of the foam associated with the drainage and the coarsening processes.\cite{mousses} In addition, the experiments are performed on short time scales, typically less than few minutes, to avoid such aging. The bubbles are generated using a $1.3\,\rm{mm}$ inner diameter needle and a syringe pump operating at constant flow rate, $Q = 20\, \mu\rm{L.min^{-1}}$. This method leads to a monodisperse foam with bubbles of diameter $D=3\,{\rm mm}$. In addition, to control the number of bubble layers, the foam is generated sequentially, one layer at a time. We also use a non-return valve to prevent the foam solution from entering in the tubing between each step.

\subsection{Oscillating forcing or impulse}

\begin{figure}
\includegraphics{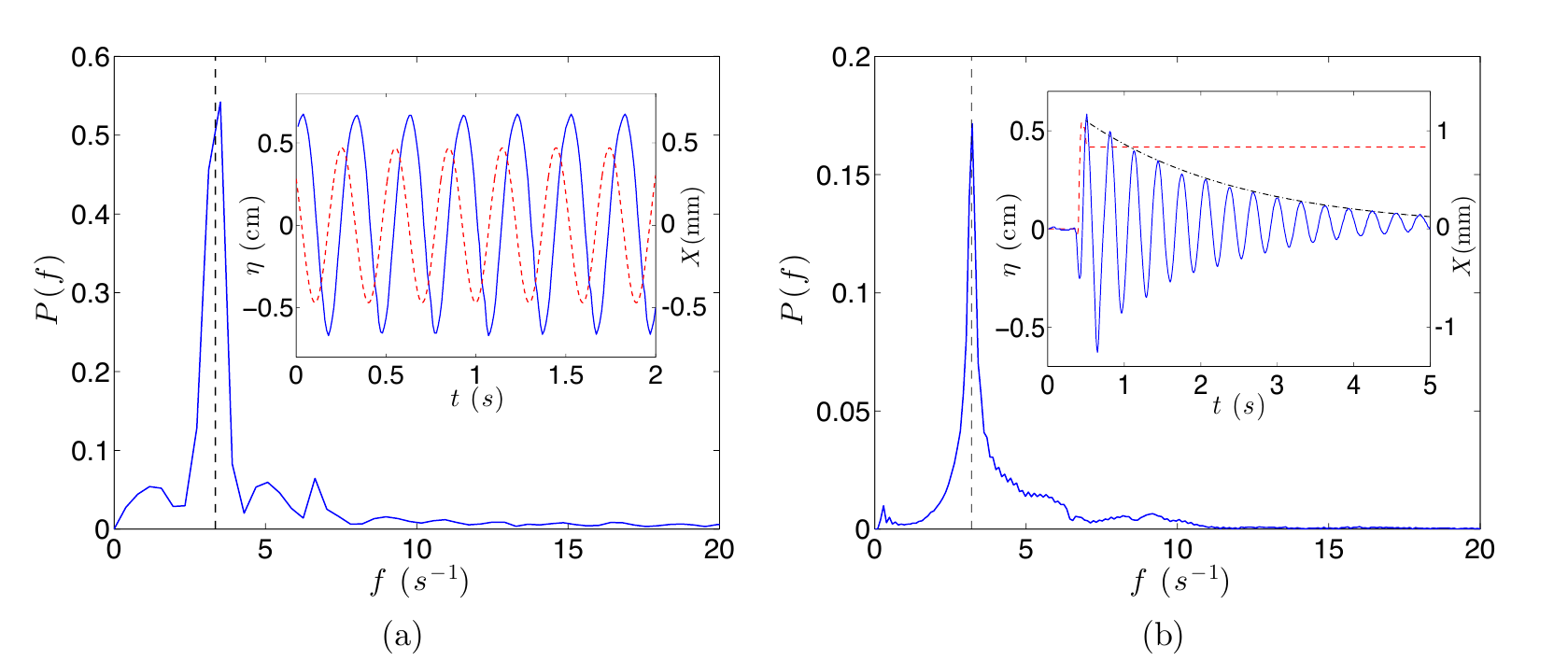}
\caption{Temporal spectrum $P(f)$ of the free-surface displacement measured $1$ cm away from the edge of the cell for (a) a continuous sinusoidal forcing at  frequency $f$=3.3 Hz and (b) an impulse excitation. The vertical dashed line shows the first resonant frequency $f_1 \simeq 3.2\,{\rm Hz}$. Insets: the corresponding time evolution of the free surface $\eta$  is represented by a blue continuous line (scale: left axis). The red dotted line shows the motion of the container $X(t)$ (scale: right axis). The black dash-dotted line in the inset of (b) corresponds to an exponential damping, $\eta_0\,\text{exp}(-t/\tau_\ell), $ where $\eta_0 $ and $ {1}/{\tau_\ell} $ are fitting parameters.  Experiments are performed with a depth $H=4\,{\rm cm}$ of foaming solution in a container of width $w=1.59\,{\rm cm}$ and length $L=7\,{\rm cm}$.}
    \label{fig:methode_impulsion}
\end{figure}

We use two methods to investigate the sloshing dynamics. In the first method, we impose a lateral harmonic excitation $X(t)=X_0\,\cos(2\pi f t)$ with a mechanical vibrator. The frequency $f$ and the amplitude $X_0$ of displacement of the moving stage are constant leading to the oscillation of the interface at the frequency $f$ as reported in figure \ref{fig:methode_impulsion}(a). This procedure allows the determination of the maximum amplitude of sloshing for different liquids, liquid heights $H$, frequencies $f$ and amplitudes of displacement of the moving stage $X_0$. In this example, a small phase shift can be observed between the harmonic lateral excitation of the container and the displacement of the free surface. This shift is likely due to the presence of nonlinear effects, which are stronger for a large container and large amplitude excitation.

In the second method, we study the response of the fluid to a mechanical impulse and record the oscillations of the interface as illustrated in figure \ref{fig:methode_impulsion}(b). We thus estimate the damping coefficient $1/\tau$ of the system. In addition, a temporal analysis of the oscillations provides the resonant frequencies of the fluid corresponding to normal modes of the container.\cite{Ibrahim}

As the aim of the present study is to determine the influence of the foam layer on the damping of the oscillations, the results presented in this article are mainly obtained using the impulse method as shown in figure \ref{fig:methode_impulsion}(b). However, determining the velocity of the bubbles appeared to be much more accurate with a forced lateral harmonic oscillation.

\noindent 


\section{Sloshing of a Newtonian fluid}\label{sec:newtonian}

The sloshing dynamics of a single Newtonian fluid resulting from a forced lateral excitation of a container has been the subject of extensive studies in various geometries such as rectangular, cylindrical or conical containers (for a review, see e.g. Ibrahim\cite{Ibrahim}). In addition to the calculation of the normal modes of the liquid in the container, the hydrodynamic pressure and the forces and moments acting on the container walls have been estimated. We should note that other forced excitations such as  pitching or rolling are also possible. In the present study, we focus on the lateral excitation of a rectangular container. Before considering the influence of the foam on the damping of the oscillations, we characterize the sloshing dynamics of a single Newtonian fluid.

\subsection{Resonant frequencies}

Assuming that the flow is incompressible, inviscid and irrotational, we use potential flow theory. The detailed derivation of the dispersion relation and velocity potential for a Newtonian fluid in a rectangular container of length $L$ and width $w \ll L$ is presented in Appendix A. The resonant frequency of order $n$, $\omega_n$, is
\begin{equation}
    {\omega_n}^{\! 2} = \left(g\,k_n+\frac{\gamma}{\rho} \,{k_n}^{\! 3}\right) \tanh(k_n\,H)
    \label{relationdispersion}
\end{equation}
where $k_n={n\pi}/{L}$ is the wavenumber, $g$ is the gravitational acceleration, $\gamma$ the surface tension and $\rho$ the density of the fluid. For a lateral time periodic forcing and a rectangular container, only the odd modes with $x$ are present in the Fourier development. Therefore, we only observe the resonant frequencies corresponding to $\omega_{2n+1}$.

\begin{figure}
    \begin{center}
\includegraphics{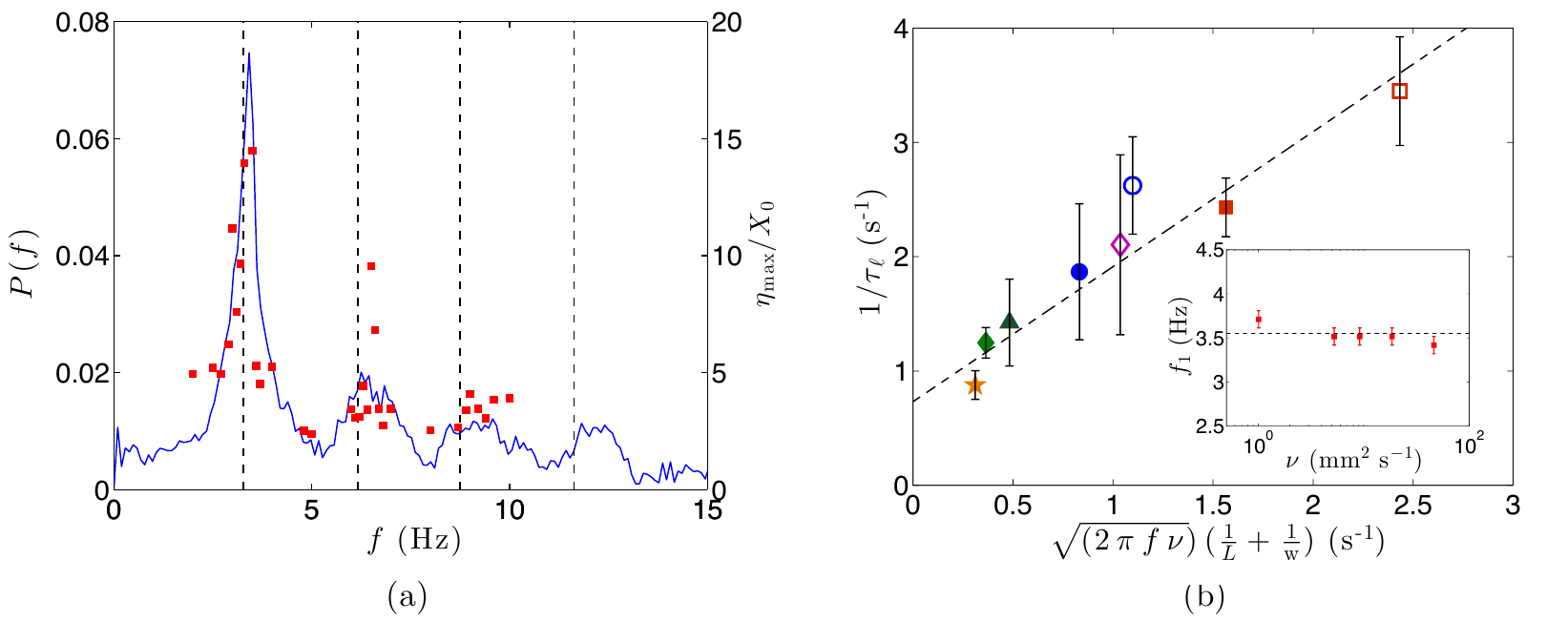}
                \caption{(a) Temporal spectrum $P(f)$ of the surface displacement resulting from an impulse excitation (blue continuous line, left axis), and with a harmonic excitation at a given frequency (red squares, right axis). The vertical dashed lines are the resonant frequencies from equation (\ref{relationdispersion}). The experiments are performed with a depth $H=4\,{\rm cm}$ of water in a container of width $w=1.59\,{\rm cm}$ and length $L=7\,{\rm cm}$. (b) Evolution of the damping coefficient ${1}/{\tau_\ell}$ for $w= 1.6$ cm and $\nu= 5.26\,\rm{mm}^2\,\rm{s}^{-1}$ (blue filled circle), $w= 1.6$ cm and $\nu= 9.17 \, \rm{mm}^2.\rm{s}^{-1}$ (blue hollow circle), $w= 1.6$ cm and $\nu= 18.6 \, \rm{mm}^2.\rm{s}^{-1}$ (red filled square), $w= 1.6$ cm and $\nu= 46.3 \, \rm{mm}^2.\rm{s}^{-1}$ (red hollow square), $w= 1.6$ cm and $\nu= 1.01 \,\rm{mm}^2.\rm{s}^{-1}$ (green filled diamond), $w= 0.48$ cm and $\nu= 1.01\, \rm{mm}^2.\rm{s}^{-1}$ (purple hollow diamond), $w= 1.1$ cm and $\nu= 1.01\, \rm{mm}^2.\rm{s}^{-1}$ (green filled triangle), $w=1.9$ and $\nu= 1.01\, \rm{mm}^2.\rm{s}^{-1}$cm (orange filled star). Inset: first resonance frequency as a function of the kinematic viscosity $\nu$.}\label{fig:relationdispersionfrequence}
    \end{center}
\end{figure}

The two experimental methods exhibit the same resonant frequency as illustrated in figure \ref{fig:relationdispersionfrequence}(a). For an impulsive motion, we record the time evolution of the free-surface elevation, $\eta(x,t)$, at a given position $x$. The resulting signal, presented in figure \ref{fig:methode_impulsion}(a), is a damped harmonic oscillation. The time series of this signal exhibit several peaks, each corresponding to a resonant frequency ($f_1$, $f_3$, $f_5$, $f_7$) as illustrated in figure  \ref{fig:relationdispersionfrequence}(a). We also observe that for the forced lateral excitation at resonant frequencies, the maximum free-surface elevation reaches local maxima. Indeed, for a harmonic forcing at a frequency $f$, the free surface oscillates at the same frequency $f$ as the excitation. Then, the amplitude of sloshing becomes large when the excitation frequency $f$ is equal to a resonant frequency of the cell ($f_1$, $f_3$, $f_5$, ...). Therefore, the results obtained with the two methods of excitation are complementary and consistent with each other.

According to the dispersion relation (\ref{relationdispersion}), the resonant frequencies depend on the height of liquid $H$ but are independent of the width of the cell $w$. The analytical expression of the resonant frequencies is consistent with the experimental results (see Appendix A). In addition, we find that for sufficient height of fluid ($k_n\,H \gg 1$), i.e. when the deep-water assumption is valid, $\omega_n$ does not depend on $H$ (see Appendix A).

Note that we also consider the influence of the viscosity on the value of the first resonant frequency measured experimentally from the impulse excitation. For the range of parameters considered, the viscosity does not influence significantly the value of $f_1$.

\subsection{Damping of the interface motion} \label{caclul_tau_1}

An important feature of the sloshing dynamics is the ability of the system to damp the waves generated by a sudden impulse excitation. The damped harmonic oscillations resulting from an impulse allow us to define the damping coefficient ${1}/{\tau_\ell}$ that depends on the physical parameters of the system: the cell width $w$, the kinematic viscosity $\nu$ and the frequency $f$. The damping coefficient is defined as \cite{gammaeau,Caps}
\begin{equation}
    \frac{1}{\tau_\ell} = \frac{ \left\langle |\dot{E_\ell} | \right\rangle}{2 E}
    \label{gamma_l}
\end{equation}

\noindent where $|\dot{E_\ell}| $ is the rate of energy dissipation in the fluid and $ E $ is the input energy. The notation $ \langle  \,\,\, \rangle$ indicates that the quantity is averaged over one period of oscillation. Experimentally, ${1}/{\tau_\ell}$ is determined using an exponential fit of the time evolution of the-free surface elevation at the position $x$: $\eta(x,t)=\eta_0 (x)\, \text{exp}(-t/\tau_\ell)\,\sin(\omega t)$. An example of an experimental fit to determine the damping coefficient ${1}/{\tau_\ell}$ is shown in figure \ref{fig:methode_impulsion}(b).
The dependence of ${1}/{\tau_\ell}$ on the parameters of the system is reported on figure \ref{fig:relationdispersionfrequence}(b). For different fluid viscosities and cell widths, we use the rescaled parameter $\sqrt{2\pi f \nu}\,(1/L+1/w)$ where $\nu$ is the kinematic viscosity of the fluid.

We assume in the modeling of the damping rate that only the first frequency of resonance contributes significantly to the damping of the sloshing dynamics for an impulse. Indeed, we did not observe a pattern of the liquid-bubble interface smaller than the cell length, which indicates that     only the first mode is dominant contrary to Faraday's instability. \cite{Caps} To estimate analytically the damping coefficient, two sources of dissipation need to be taken into account.\cite{Landau1959} The first one corresponds to viscous forces in the bulk of the fluid:
\begin{equation}
   \left \langle | \dot{E_\ell} | \right\rangle_1= \frac{1}{2} \rho \, \nu \int_{\rm bulk} {\left(\bm{\nabla} \bm{v} + \bm{\nabla} \bm{v}^T \right)^2 \rm{d}V} \sim \rho \, \nu \left(v k\right)^2 \,\frac{L\,{w} }{k}
    \label{eq:puissance1}
\end{equation}
The term proportional to $1/k$ arises from the sum of the velocities along the $z$-axis in the deep water regime. The second source of damping comes from the viscous effects in the boundary layers, namely the oscillating Stokes layers of thickness $\delta=\sqrt{{\nu}/{\omega}}$ near the two walls:\cite{Caps}
\begin{equation}
    \left \langle | \dot{E_\ell} | \right\rangle_2 = \frac{1}{2} \rho \, \nu \int_{\rm BL} {\left(\bm{\nabla} \bm{v} + \bm{\nabla} \bm{v}^T  \right)^2 \rm{d}V}  \sim \rho \, \nu \,\left(\frac{v}{\delta}\right)^2  \left(L+w\right) \frac{\delta}{k}
    \label{eq:puissance2}
\end{equation}

\noindent Considering $k = k_1 = \pi/L$ and $w\ll L$, the ratio of equations (\ref{eq:puissance1}) and (\ref{eq:puissance2}) is typically of the order of $\pi \delta w /L \sim 10^{-2}$ in our system. Therefore, the damping in the bulk is negligible compared to the contribution in the wall boundary layers in our experiments.
Consequently, the dissipated power can be estimated as:
\begin{equation}
    \left \langle \dot{|E_\ell|} \right \rangle \sim \rho \, \nu \,\left(\frac{v}{\delta}\right)^2  \left(L+w\right) \frac{\delta}{ k}\label{eq:El_dissipated}
\end{equation}
The initial energy injected in the system is equal to the initial kinetic energy and is defined as \cite{Landau1959}
\begin{equation}
    E = \frac{1}{2} \rho \int {\bm{v}^2 \rm{d}V} \sim \frac{1}{2}\,\rho \, v^2  \frac{L\, {w}}{2\, k}\label{eq:El_injected}
\end{equation}
for which the integral is taken over the volume $w\,L/(2\,k)$. Substituting equation (\ref{eq:El_dissipated}) and (\ref{eq:El_injected}) in the definition (\ref{gamma_l}) of the damping coefficient ${1}/{\tau_\ell}$ leads to
\begin{equation}
    \frac{1}{\tau_\ell} \sim \, \sqrt{2\pi\, f\,\nu} \left(\frac{1}{L} + \frac{1}{{w}}\right).
    \label{energiecin}
\end{equation}

We observe on the figure \ref{fig:relationdispersionfrequence}(b) that this scaling law is in good agreement with the experimental measurements for various fluids and cell thicknesses. These results confirm that the dissipation in the wall boundary layer is the main source of damping of the sloshing in our system.


\section{Sloshing damped by a liquid foam}\label{sec:foam}

We now consider the influence of a liquid foam placed on top of a vibrating liquid bath.\cite{Cappello2014} The experiments presented in this section were performed with a liquid height $H=4$ cm of foaming solution in a container of length $L=7\,{\rm cm}$. The liquid foam is made of monodisperse bubbles of diameter $D=3$ mm and the height of the foam layer is $\xi D$ where $\xi$ represents the number of layers.

\subsection{Experimental observations in a Hele-Shaw cell}

To allow visualization of the motion of the bubbles, we first perform experiments in a confined cell of width $w=3$ mm equal to the bubble diameter $D$. The foam is generated by injecting air at a constant flow rate through a needle located at the bottom of the cell. In all experiments, no topological rearrangements (T1) are observed.\cite{Weaire1988}
As a result, the foam behaves like a visco-elastic fluid without plastic deformation.

\begin{figure}
    \centering
\includegraphics{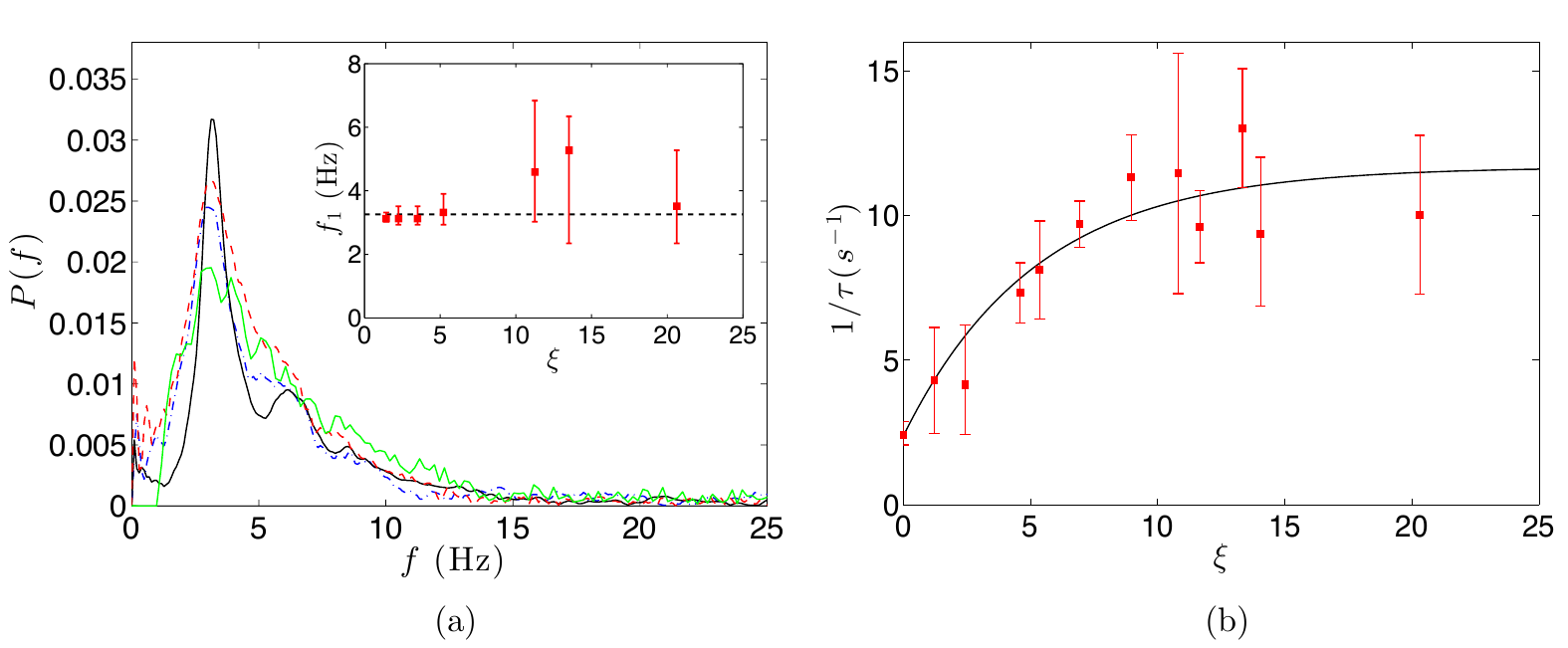}
    \caption{(a) Influence of the foam on the temporal spectrum of the fluid response to an impulse excitation motion $P(f)$ for the foam solution alone (black continuous line) or with one (red dashed line), two (blue dashed-dotted line) and three (light green continuous line) layers of bubbles. Inset: evolution of the first resonant frequency $f_1$ as a function of the number of layers of bubbles $\xi$, the horizontal dashed line shows the theoretical frequency. (b) Evolution of the damping coefficient $1/\tau$ as a function of the number of layers $\xi$. The black continuous line shows the fit by equation (\ref{eq:fitgamma}). Both figures are obtained with experiments performed with a depth $H=4\,{\rm cm}$ of foaming solution in a container of width $w=0.30\,{\rm cm}$ and length $L=7\,{\rm cm}$.}
    \label{fig:gammafoam}\label{fig:tfmousse}
\end{figure}

Experimentally, we consider the influence of the foam on the resonant frequency of the dominant mode of sloshing, i.e. the mode corresponding to the frequency $f_1$.
Figure \ref{fig:tfmousse}(a) shows the temporal spectrum extracted from the oscillations of the liquid-foam interface following an impulse motion.
In the presence of a foam, the spectrum exhibits a peak located around $f_1 \simeq 3.1$ Hz.
This value is similar to the frequency of the first mode in the absence of foam, which is predicted to be $3.3$ Hz from equation (\ref{relationdispersion}).
This result suggests that the presence of foam does not significantly modify the characteristic frequency of the system. To characterize the foam thickness, we introduce $\xi$, the number of layers of bubbles, as the ratio of the foam height and the bubble diameter.
The resonant frequency $f_1$ is measured experimentally for different foam thicknesses, i.e. values of $\xi$ (see the inset of figure \ref{fig:tfmousse}(a)). As no significant change is observed, the frequency of the first mode is assumed constant.
This result indicates that the elasticity at the interface between the air and the foam can be neglected at first order thus leading to no supplementary contribution in the dispersion relation (\ref{relationdispersion}).\cite{Planchette2012}

\begin{figure}
    \centering
    \includegraphics{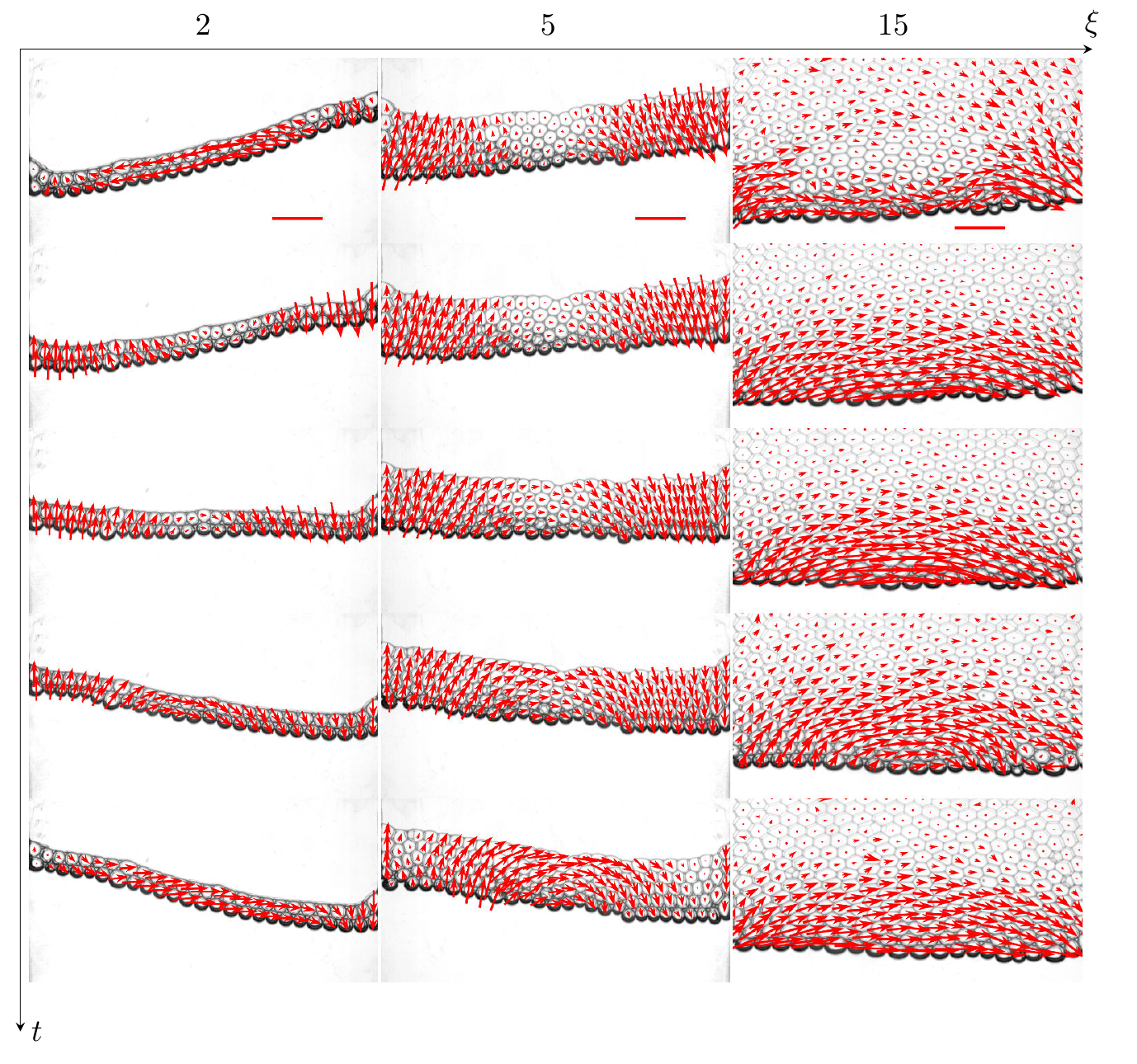}
    \caption{Time evolution of the amplitude of sloshing for a harmonic lateral forcing at $f_1=3.125$ Hz (first resonant frequency) and an amplitude of excitation $X_0=0.1$ cm. The number of layers of bubbles $\xi$ increases from left to right. The arrows indicate the velocity field. The photographs are taken at different moments during half a period of the container oscillation. Scale bars are $1$ cm. (Multimedia View).}\label{fig:timelaps}
\end{figure}

For a continuous harmonic forcing at the resonant frequency $f_1$ and lateral displacement amplitude $X_0=1\,{\rm mm}$, the maximum amplitude of sloshing decreases significantly for increasing foam thicknesses as shown in figure \ref{fig:timelaps} (Multimedia View).
    Indeed, without foam, the free-surface has a maximum amplitude of the order of $1$ cm; the addition of only $\xi=5$ layers of bubbles leads to a maximum amplitude of the order $1$ mm, which is a factor of ten reduction. To evaluate the damping coefficient for various foam thicknesses, we perform experiments with an impulse motion and measure the time evolution of the oscillations of the liquid-foam interface (see figure \ref{fig:methode_impulsion}(b)). The experimental results reported in figure \ref{fig:tfmousse}(b) show that the damping coefficient $1/\tau$ increases exponentially toward a plateau over a typical rescaled length corresponding to $5$ layers of bubbles. This result suggests that the addition of a few layers of bubbles is sufficient to efficiently damp the sloshing. These qualitative observations confirm that foam has a strong damping effect on the sloshing dynamics, decreasing the amplitude of the oscillations and increasing the damping coefficient of the system $1/\tau$.

To understand the effect of the foam, the damping coefficient is decomposed into two contributions.
The first contribution ${1}/{\tau_\ell}$ is due to the liquid alone, and depends only on the frequency $f$, the kinematic viscosity $\nu$ and the size of the cell.
The second contribution to damping $1/\tau_f$, is induced by the presence of the foam.
Thus, we assume $1/\tau={1}/{\tau_\ell}+{1}/{\tau_f}$.
It is noteworthy that $\tau_\ell$ is independent of the position or velocity of the interface. This makes the decomposition valid even in presence of the foam that reduces the amplitude of the liquid-foam interface. Then, the experimental values of the damping coefficient of the system can be fit by:

\begin{equation}
    \frac{1}{\tau} = \frac{1}{\tau_\ell}+\frac{1}{\tau_f^{\infty}}\,\left[1-\text{exp}\left(-\frac{\xi}{\xi_c}\right)\right]
    \label{eq:fitgamma}
\end{equation}

\noindent as illustrated in figure \ref{fig:tfmousse}(b) where $1/\tau_f^{\infty}=9\pm3\,\,\rm{s^{-1}}$, a plateau ${1}/{\tau_\ell} + 1/\tau_f^{\infty}=11\pm3 \, \rm{s^{-1}}$ and $\xi_c=5\pm 2$ for the considered experimental parameters. The damping coefficient of the foam solution alone is estimated to be ${1}/{\tau_\ell} \simeq 2 \,\rm{s^{-1}}$, which is in agreement with the results presented in figure \ref{fig:relationdispersionfrequence}(b) for a height of foaming solution $H=4$ cm in a cell of width $w=3$ mm and length $L=7$ cm. These experimental results show that the foam-air interface is substantially stabilized with a sufficient amount of foam. 

We have also recorded the dynamics for a harmonic forcing for various numbers of foam layers $\xi$.
At sufficiently large foam thicknesses, $\xi > 5$, the damping coefficient $1/\tau$ and the maximum free-surface elevation $\eta$ reach a plateau value. This saturation is consistent with the observations reported above: when the thickness of the foam is sufficiently large, typically $\xi > 5$, the foam-air interface remains nearly steady.

Also, we have measured the velocity of individual bubbles with bubble labeling\cite{Vanderwalt2014} and particle tracking techniques\cite{Daniel2014} as shown in figure \ref{fig:timelaps} (Multimedia View). We observe that adding a large number of layers of bubbles allows to document the spatial dependence of the velocity of the bubbles. For fewer than two layers of bubbles the velocity is mostly vertical.
The velocity becomes more horizontal as the number of layers increases, except for bubbles near the vertical walls, which have a non-negligible vertical velocity due to the boundary conditions.
As represented in figure \ref{fig:gamma2final}(a), we can observe that the mean bubble velocity decreases exponentially $ \langle |v_b| \rangle \propto\text{exp}(-z/(\xi_c D) )$ with a characteristic length scale of $\xi_c D = 16.8 \pm 1.8\,{\rm mm}$ corresponding to $\xi_c\simeq5$ layers of bubbles.
We also noted that the mean velocity of the interface is independent of the $x$ position.


\subsection{Phenomenological modeling}

A full analytical description of sloshing dynamics in presence of the foam is beyond the scope of this study as the physical description of the motion of the foam remains mainly empirical. Instead, in this section we rationalize the evolution of the damping coefficient $1/\tau$ presented in figure \ref{fig:gammafoam}. To do so, we describe the contribution of the addition of foam to the damping coefficient using a phenomenological model. Here, our approach remains the same as the calculation of the damping coefficient for a single Newtonian fluid $1/\tau_{\ell}$ presented in section \ref{caclul_tau_1}. We consider a $2$D foam confined in a cell of width $w$.

 The foaming solution used in the experiments leads to a foam with a low surface modulus.\cite{Denkov2009}
 Thus, the friction force of the Plateau borders on two solid walls can be estimated as $f_b = K\, \gamma_f \, {\rm Ca}^{2/3}$, where ${\rm Ca} = \mu_f\,v_b/\gamma_f$ is the Capillary number and $K$ is a constant for given geometrical parameters describing the Plateau borders.\cite{Raufaste2009,Cantat2013}
The typical Capillary number considered in our experiments is of order ${\rm Ca} \sim  3 \times 10^{-3}$, where $\mu_f$ and $\gamma_f$ are the viscosity and surface tension of the foaming solution, respectively. Note that the present scaling for the friction force is no longer valid for ${\rm Ca} > 10^{-2}$ due to inertial effects, but our experiments remain below this limit.\cite{Quere1998}

We denote $v_b(x,z,t)$, the velocity of the bubble located at $(x,z)$. The rate of energy dissipation by the foam, averaged over a period, can be written:
\begin{equation}
\left\langle | \dot{E_f} | \right\rangle = \left\langle \sum f_b\,v_b  \right\rangle,
\end{equation}
where the sum is taken over all bubbles constituting the foam.
The shape of the Plateau borders is assumed to be independent of $x$ (translational symmetry) and $z$, i.e., we neglect the variation of the liquid fraction along the vertical direction, such that $K$ is independent of $z$.
Upon substitution, we obtain
\begin{equation}    \label{puissance}
   \left\langle | \dot{E_f} | \right\rangle  =K'\, \sum {\left\langle {v_b}^{5/3} \right\rangle}
\end{equation}
with $K'= K {\gamma_f}^{1/3}\,{\mu_f}^{2/3}$.

\begin{figure}
    \centering
\includegraphics{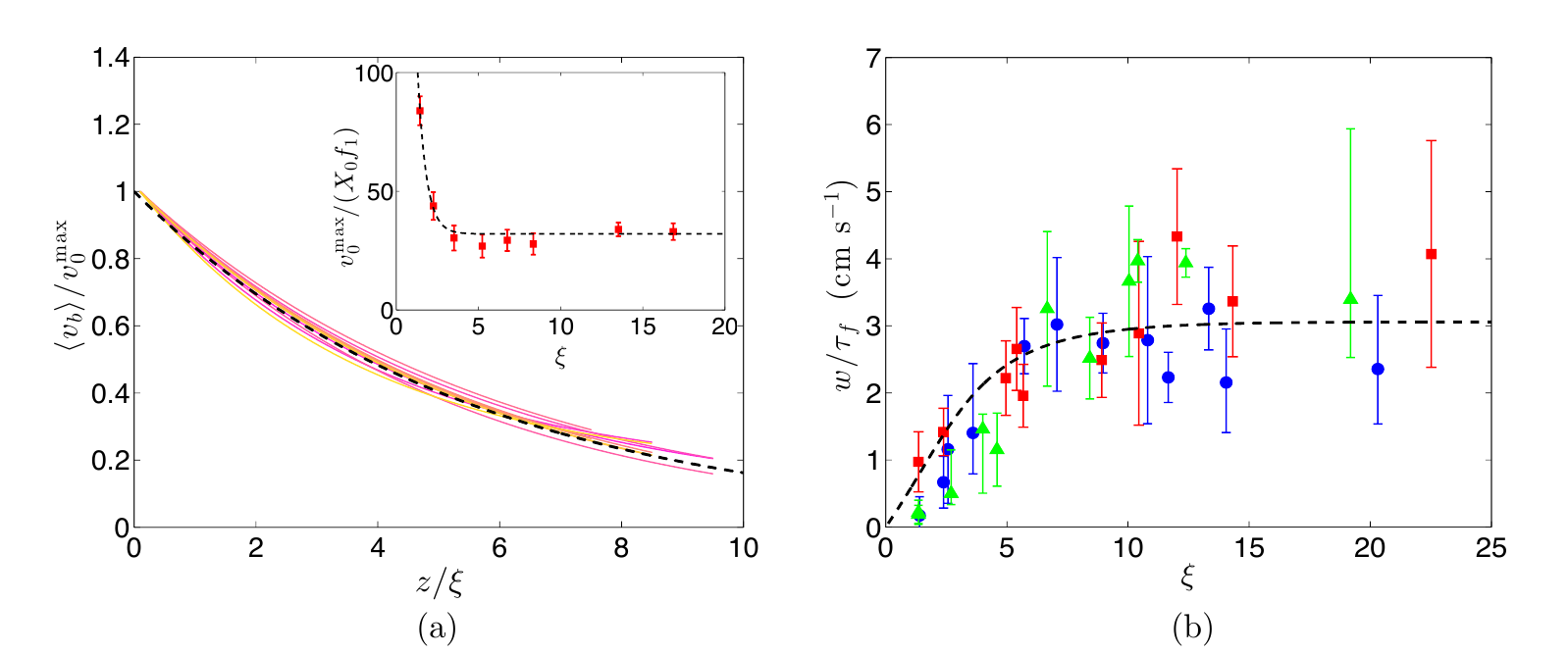}
    \caption{(a) Velocity of the bubbles $v_b$ averaged over one period and normalized by $v_0^{max} = v_b(z/\xi=0)$ as a function of the vertical position $z/\xi$ in a foam made of $9$ layers of bubbles. The curves correspond to different locations $x$ and the black dotted line is the exponential fit $\exp[-z/(5.2\,\xi)]$. Inset: Evolution of the rescaled maximum value of the mean velocity of the bubbles $v_0^{max}/(X_0\,f_1)$ as a function of $\xi$. The dotted line corresponds to an exponential fit of the experimental measurements. These measurements are obtained in a $w=0.3\,{\rm cm}$ wide cell of length $L=7\,{\rm cm}$ and a height of foaming solution $H=4\,{\rm cm}$. (b) Rescaled damping coefficient $w/\tau_f$ as a function of $\xi$ for various widths of the cell: $w=0.3$ cm (blue circles), $w=0.6$ cm (green triangles) and $w=1.6$ cm (red squares) in $L=7\,{\rm cm}$ cells. The black dashed line shows the best fit adjusting the prefactor in the equation (\ref{eq:formuletheorique}).}\label{fig:gamma2final}
\end{figure}

Experimentally, we observe that the time-averaged velocity $\langle v_b \rangle$ does not depend significantly on the horizontal position of the bubble $x$.
In addition, as noted previously, the velocity of the bubbles can be fit by an exponential curve $v_b(z,t)=v_0^{max}\,\text{exp}(-z/(\xi_c D))\,\cos(\omega\,t)$ where $\xi_c D = 16.8 \pm 1.8\,{\rm mm}$ as illustrated in figure \ref{fig:gamma2final}(a).
We note that the order of magnitude of the lengthscale over which the velocity decreases has the same order of magnitude as the length scale associated with the first mode of resonance $1/k_1=L/\pi \simeq 22\,{\rm mm}$.
The maximum velocity $v_0^{max}$ is obtained at $z=0$, i.e. for the first bubble layer on top of the liquid bath (see inset of figure \ref{fig:gamma2final}(a)).

Since we have assumed that the velocity of the bubbles is independent of the horizontal position, the integral over the Plateau border in the $x$-direction leads to a prefactor $L/D$, which corresponds to the number of bubbles.
Thus, equation (\ref{puissance}) becomes
\begin{equation}
    \left\langle | \dot{E_f} | \right\rangle  ={K'} \frac{L}{D} \int \limits_{0}^{\xi D}\, \left\langle (v_0^{max})^{5/3}\,|\text{cos}(\omega  t)|^{5/3} \right\rangle \,
    \text{exp}\left(-\frac{5\,z}{3\,\xi_c D} \right)\, {\rm d}z
\end{equation}
where $\xi D$ is the height of foam.
Therefore, the dissipated power can be approximated as
\begin{equation}
    \langle | \dot{E_f} | \rangle  =\frac{3}{5} \xi_c\, L \,{K'}\,(v_0^{max})^{5/3} \left\langle |\text{cos}(\omega t)|^{5/3} \right\rangle  \left[1-\text{exp}\left(-\frac{5\,\xi}{3\,\xi_c} \right)\right].\label{eq:Edot}
\end{equation}

The mechanical energy injected initially in the system can be approximated as the maximum of the kinetic energy of the fluid, which is reached when the liquid-foam interface is horizontal. Therefore, the injected energy is:
\begin{equation}
    E = \frac{1}{2}\rho\int{(v_{0}^{max})^{2} \rm{d}V} = \frac{\rho \, {w}}{4}\, \frac{L^2}{\pi}(v_{0}^{max})^{2}.  \label{eq:E}
\end{equation}
Finally, combining equations (\ref{eq:Edot}) and (\ref{eq:E}), we obtain the damping coefficient of the foam:
\begin{equation}
    \frac{1}{\tau_{f} } = \frac{12 \pi}{5} \frac{\xi_c\,{K'}  \left\langle |\cos(\omega t)|^{5/3} \right\rangle\,}{  \rho\, w\, {L} (v_{0}^{max})^{1/3} }  \left[1-\text{exp}\left(-\frac{5\,\xi}{3\,\xi_c}     \right)\right].
    \label{eq:formuletheorique}
\end{equation}

The value of the maximum velocity $v_{0}^{max}$ is extracted from experimental measurements (see figure \ref{fig:gamma2final}(a)).
Substituting $(v_{0}^{max})^{1/3}$ with its experimental fit, we can estimate the evolution of $1/\tau_f$. The phenomenological law is reported in figure \ref{fig:gamma2final}(b) and shows a fairly good agreement with the experimental measurements for different container sizes and different foam thicknesses.

The prefactor is determined by fitting the experimental data to find
\begin{equation}
\frac{12 \pi}{5} \frac{\xi_c\,{K'}  \left\langle |\cos(\omega t)|^{5/3} \right\rangle\,}{  \rho\, {L} (v_{0}^{max})^{1/3} } = 2.76 \,\textrm{cm/s}
\end{equation}
which leads to $K \approx 1.9 $ m/s with $v_{0}^{max}\approx 0.1$ m/s.
The value of $K$ can be compared with the phenomenological law\cite{Cantat2013}
\begin{equation}
K = 5.13 \left(\frac{R_p}{\sqrt{A}}\right)^{-0.48}\label{eq:Cantat}
\end{equation}
where $R_p$ is the radius of curvature of the Plateau border and $A$ is the bubble area.

In the present study, we assumed that $K$ does not depend on the liquid fraction as its measurement during the oscillations is difficult.
In addition, the strongest variation of the liquid fraction happens between the first and the second layer of bubbles because the first layer is directly in contact with the liquid bath (see figure \ref{fig:timelaps}).
The change in liquid fraction is less significant between the 2nd and the 5th layers, where most of the dissipation occurs according to the velocity profiles.
Thus, we estimate the ratio ${R_p}/{\sqrt{A}}$ to be of the order of $0.3$.
Note that the dependency on this ratio is only weak because of the power law in equation (\ref{eq:Cantat}).
Consequently, with this order of magnitude estimate, we obtain a prediction of $K=9.1$ m/s.
These two values are compatible, especially as it is not possible to estimate the bubble shape accurately.


\subsection{From 2D to confined 3D foams}

We performed experiments with an impulse excitation for three values of the cell thickness $w$: $3$, $6$ and $15.9$ mm which corresponds to $1$, $2$ and $5$ bubbles diameters in the $z$ direction, respectively. We follow the experimental procedure developed for the 2D foam to determine the damping coefficient. We describe these foams as confined 3D foams since we consider that the sloshing of the interface can be described by the two-dimensional modeling.

The phenomenological model developed in the previous section relies on an hypothesis: the dissipation that enhances damping is only induced by the viscous friction between the bubbles and the walls of the container. To consider a confined $3$D foam, we consider that only the Plateau borders in contact with the walls contribute to the energy dissipation and that the bubbles in the bulk are simply advected by the bubbles in contact with the walls. Thus, neglecting the contribution to the damping arising from the bulk, the damping coefficient, $1/\tau_f$, should scale as the inverse of the width of the cell, $1/w$.

The evolution of the damping coefficient for a $3$D confined foam is reported in figure \ref{fig:gamma2final}(b) as a function of the number of layers $\xi$ of bubbles. The results for the different width collapse well on a master curve.
In addition, the experimental results can be fit with the exponential behavior obtained in the equation (\ref{eq:formuletheorique}).
The exponential trend leads to a value of the plateau ${w}/\tau_f^{\infty}=1.6 \pm 0.3 \,\, \rm{cm}\,\rm{s}^{-1}$ whereas the best fit suggests that the plateau has a value of ${w}/\tau_f^{\infty} \simeq 3.1 \,\, \rm{cm}\,\rm{s}^{-1}$. The phenomenological model gives a good order of magnitude of the contribution of the foam to the damping coefficient. The relative error in prefactor can likely be explained by the shear between the bubbles which can induce some dissipation. We also observe in figure \ref{fig:gamma2final}(b) that the saturation value of the rescaled damping coefficient $w/\tau_f$ seems to slightly increase with the width of the cell. This observation indicates that the dissipation in the bulk could add a small contribution to the damping for a $3$D foam.

These results and the phenomenological model confirm that the foam is responsible for the damping of the free-surface oscillations. A model based on the viscous dissipation of the foam against the walls of the container captures the evolution and the order of magnitude of the damping coefficient in 2D and confined 3D situations. It is important to stress that the cells used in this study are narrow, which means that the number of bubbles in the transverse direction does not exceed 6 bubbles.
Thus, further investigation of 3D foams is necessary to fully characterize the role of the third dimension. In addition, to refine the present model, the influence of the foam elasticity as well as the dissipation in the bulk should be taken into account.

\section{Conclusion}\label{sec:conclusion}

In this paper, we have studied how the presence of foam on top of a sloshing liquid increases the damping coefficient and reduces the amplitude of the free-surface oscillations. First, we characterize the sloshing dynamics in the linear regime for a Newtonian fluid in Hele-Shaw cells. We show that the measured resonant frequencies are well described by a wave model based on the inviscid and irrotational flow hypothesis. Moreover, the damping coefficient of the interface motion can be attributed to the fluid viscosity through the oscillating Stokes boundary layers on the two walls.

Then, the influence of the foam thickness is experimentally quantified for 2D and confined 3D foams.
We report that the resonant frequencies are not modified significantly by the presence of foam. We show that the damping effects of the foam is significant even with only $5$ layers of bubbles. For a larger number of layers, the damping coefficient and the amplitude of the oscillations become independent of the foam thickness. For both 2D and confined 3D foams, we rationalize our experimental results using a phenomenological model that accounts for the viscous dissipation induced by the motion of Plateau borders against the walls.

This study demonstrates that a relatively thin layer of foam effectively damps sloshing. Our findings suggest that foam could be used  in various industrial processes in which sloshing needs to be minimized. Future projects should consider additional effects and parameters relevant to those applications. For instance, when the width of the container becomes comparable to the height of foam, the interface motion becomes truly three dimensional and the contribution of the bulk to the dissipation need to be taken into account. Other parameters can be considered to enhance the damping effect such as the interfacial rigidity that depends on the physicochemical properties of the surfactants\cite{dressaire2008} and the roughness of the walls that could increase the friction.

\acknowledgments

The authors thank Isabelle Cantat, Marie-Caroline Jullien and Jonathan Katz for helpful discussions and comments. This research was made possible in part by the CMEDS grant from BP / The Gulf of Mexico Research Initiative.

\appendix

\section*{Appendix A: Sloshing of a single Newtonian fluid}

\subsection{Analytical potential model}

\begin{figure}
    \centering
  \includegraphics{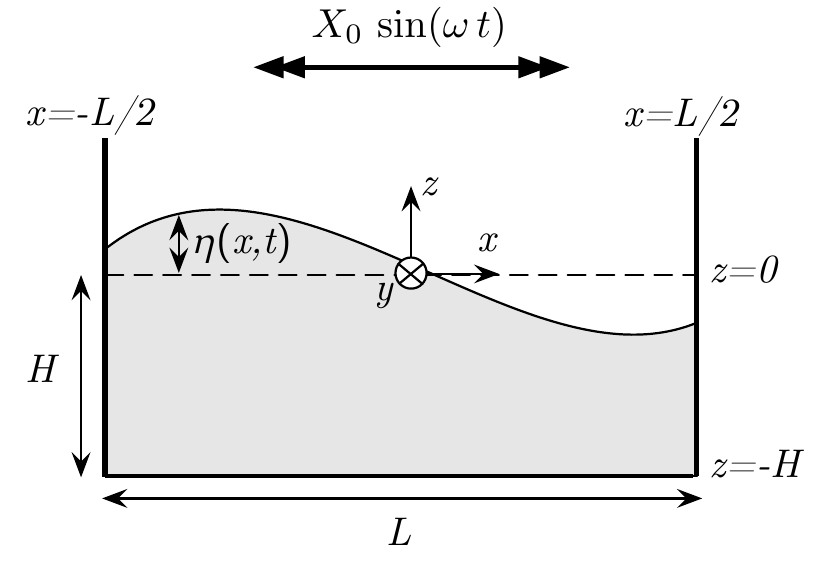}
    \caption{Two-dimensional schematic of the sloshing dynamics in a rectangular container.}\label{fig:appendix}
\end{figure}

In the following we consider the small lateral excitation $X(t)=X_0\,\sin(\omega\,t)$ of a rectangular rigid container of length $L$ and width $w \ll L$ such that the motion of the fluid remains two-dimensional (figure \ref{fig:appendix}). The height of the liquid at rest is $H$. The theoretical analysis of the response of the fluid relies on the inviscid, incompressible and irrotational flow assumptions. In this framework, the velocity field of the fluid derives from a potential $\Phi$ such that $\bm{v}=- \nabla \Phi$. The potential $\Phi$ satisfies the Laplacian equation:
\begin{equation}
    \nabla^2 \Phi=0
    \label{appendix_laplacien}
\end{equation}

The potential $\Phi$ is the sum of the potential $\Phi_0$ associated to the oscillating container, $\Phi_{0}=-X_0\,\omega\,x\,\cos(\omega\,t)$ and the potential $\tilde{\Phi}$ that describes the relative motion of the fluid in the reference frame of the container such that $\Phi=\Phi_{0}+\tilde{\Phi}$. In the frame of reference attached to the container, the boundary conditions on the bottom $z=-H$ and the sidewalls $x=\pm{L}/{2}$ are
\begin{equation}
\frac{\partial \tilde{\Phi}}{\partial z}=0\quad \text{at} \quad z=-H \qquad \text{and} \qquad  \frac{\partial \tilde{\Phi}}{\partial x}=0\quad \text{at} \quad x=\pm\frac{L}{2}.\label{appendix_condition_1}
\end{equation}

\noindent The linearized kinematic and dynamic conditions at the free interface $\eta(x,t)$ lead respectively to \cite{Ibrahim,LaRocca}
\begin{eqnarray}
    -\frac{\partial \tilde{\Phi}}{\partial z}=\frac{\partial \eta}{\partial t} \qquad \text{at} \qquad z=\eta(x,t),   \label{vitesse} \\
  {\rm and} \qquad\qquad  g\eta-\frac{\partial \tilde{\Phi}}{\partial t}+\frac{\gamma}{\rho}\kappa+\ddot{X}(t)\,x=0 \qquad \text{ at} \qquad z=\eta(x,t), \label{bernoulli}
\end{eqnarray}
where dots indicate time derivatives, $\eta(x,t)$ denotes the position of the free interface measured from its equilibrium position, $g$ is the gravitational acceleration, $\gamma$ is the surface tension and $\rho$ is the density of the fluid. In the linear assumption, the curvature of the free interface is $\kappa=-\partial^2 \eta/\partial x^2$.

The solution of the equation (\ref{appendix_laplacien}) with the boundary conditions (\ref{appendix_condition_1})-(\ref{bernoulli}) is given by:
\begin{equation}
    \Phi(x,z,t)=-X_0\,\omega\, \cos(\omega t)\left[x+\frac{4}{L}\,\sum_{n=0}^{\infty}\left(\frac{(-1)^{n}\sin(k_{2n+1}x)\cosh(k_{2n+1}z)}{{k_{2n+1}}^2 \cosh(k_{2n+1}h)}\frac{\omega^2}{({\omega_{2n+1}}^2-\omega^2)}\right)\right]
    \label{Phi}
\end{equation}

\noindent where $\omega=2\,\pi\,f$ and $X_0$ are the angular frequency and the amplitude of oscillation of the moving stage, respectively. In addition, $k_n={n\pi}/{L}$ is the wavenumber and $\omega_n$ is the fluid-free-surface natural frequency. Using the relation (\ref{bernoulli}) we obtain

\begin{equation}
    \eta(x,t)=\frac{X_0}{g}\,\omega^2 \, \sin(\omega t) \left[ x+\sum_{n=0}^{\infty} \left(\frac{(-1)^{n}\sin(k_{2n+1}x)}{L\,{k_{2n+1}}^2}\frac{\omega^2}{({\omega_{2n+1}}^2-\omega^2)}\right)\right]
    \label{eta}
\end{equation}

\noindent The resonant frequency of order $n$ of the system, $\omega_n$, satisfies
\begin{equation}
    {\omega_n}^{\! 2} = \left(g\,k_n+\frac{\gamma}{\rho} \,{k_n}^{\! 3}\right) \tanh(k_n\,H)
    \label{appendix_relationdispersion}
\end{equation}

We notice that only the odd modes with respect to $x$ are present in the Fourier developments for $\Phi$ (see equation \ref{Phi}) and $\eta$ (see equation \ref{eta}); the $x$-component of the velocity field vanishes at $x=\pm L/2$, i.e. $\partial \tilde{\Phi}/\partial x=0$ at $x=\pm L/2$.

\subsection{Experimental characterization}

\begin{figure}
    \begin{center}
\includegraphics{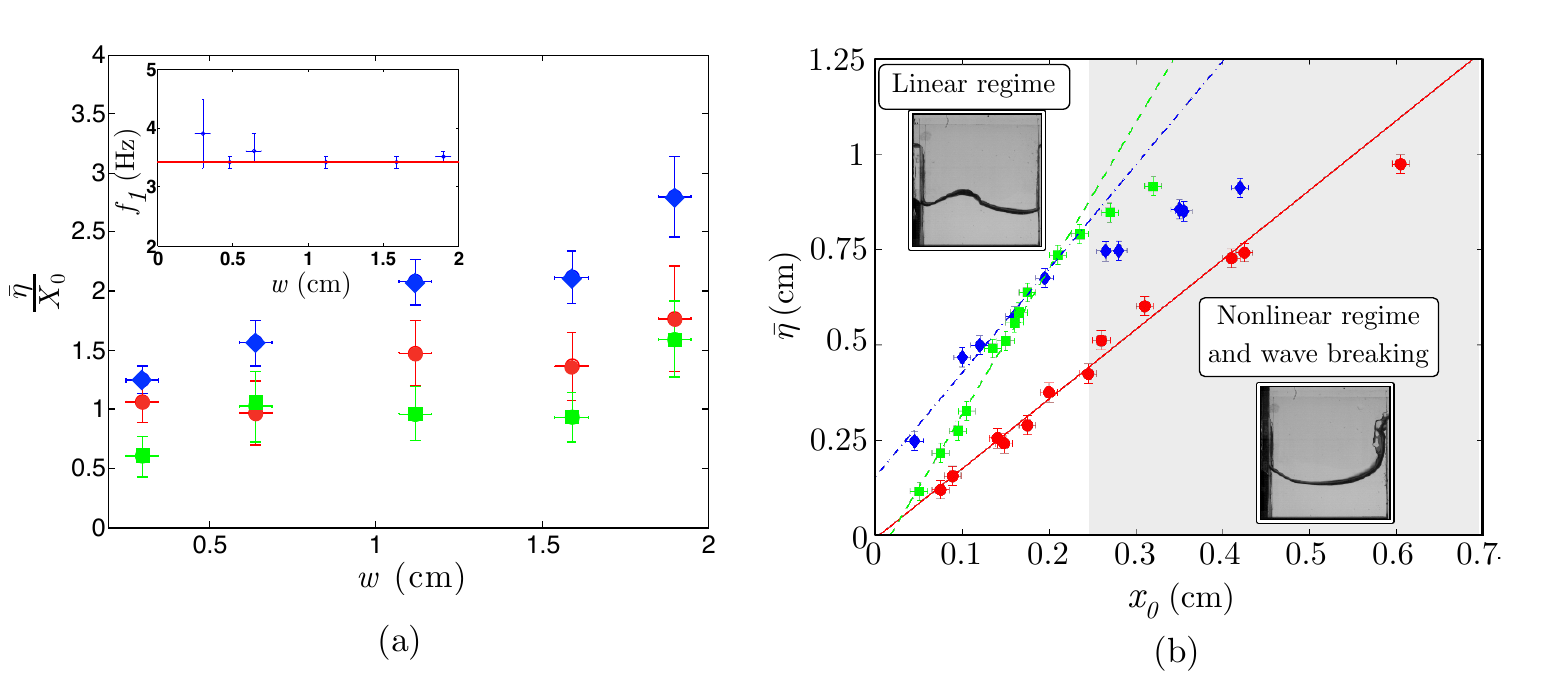}
        \caption{(a) Rescaled maximum amplitude $\bar{\eta}/{X_0}$ averaged over $x$ as a function of the width of the cell $w$ for a forcing frequency $f=4$ Hz (blue diamonds), $f=5$ Hz (red circles) and $f=7$ Hz (green squares). Inset: evolution of the first resonant frequency $f_1$ as a function of $w$. (b) Maximum amplitude of sloshing $\bar{\eta}$ averaged over $x$ as a function of the amplitude of the moving stage $X_0$ for a frequency $f= 3.1$ Hz and a width $w= 0.3$ cm (green squares), a frequency $f= 3.5$ Hz and a width $w=1.6$ cm (blue diamonds) and a frequency $f=4.8$ Hz and a width $w=1.6$ cm (red circles).}
        \label{fig:influencegap}
    \end{center}
\end{figure}

We first characterize the effect of the width of the cell on the amplitude of sloshing and resonant frequencies using a Newtonian fluid, water. We set a harmonic forcing $X(t)=X_0\,\cos(\omega\,t)$ and measure the maximum amplitude of sloshing along the $x$-axis for various forcing frequencies and cell widths. The results are reported in figure \ref{fig:influencegap}(a).

Experiments with a Newtonian fluids are performed with cells of width $w \in[3; 15.9]$ mm, which are sufficiently thin to avoid three-dimensional effects and sufficiently large to separate the boundary layer effects from the flow in the bulk. The dispersion relation (\ref{appendix_relationdispersion}) shows that the resonant frequencies do not depend on the width $w$ of the cell. This result is confirmed experimentally in the inset of figure \ref{fig:influencegap}(a). Studies with two-dimensional foam are performed with a cell of width $w = 3$ mm to remain in the regime where the potential theory can be used while allowing the visualization of the bubbles.

The analysis presented in the previous section assumes that the response of the fluid remains within the linearized dynamics. In this study we defined a range of experimental parameters such that the fluid motions are in the linear regime as shown in the left part of figure \ref{fig:influencegap}(b). In the following, we ensure that the harmonic excitation or  the impulse forcing are sufficiently low to avoid wave breaking which would add complexity to the analysis.

\begin{figure}
    \begin{center}
     \includegraphics{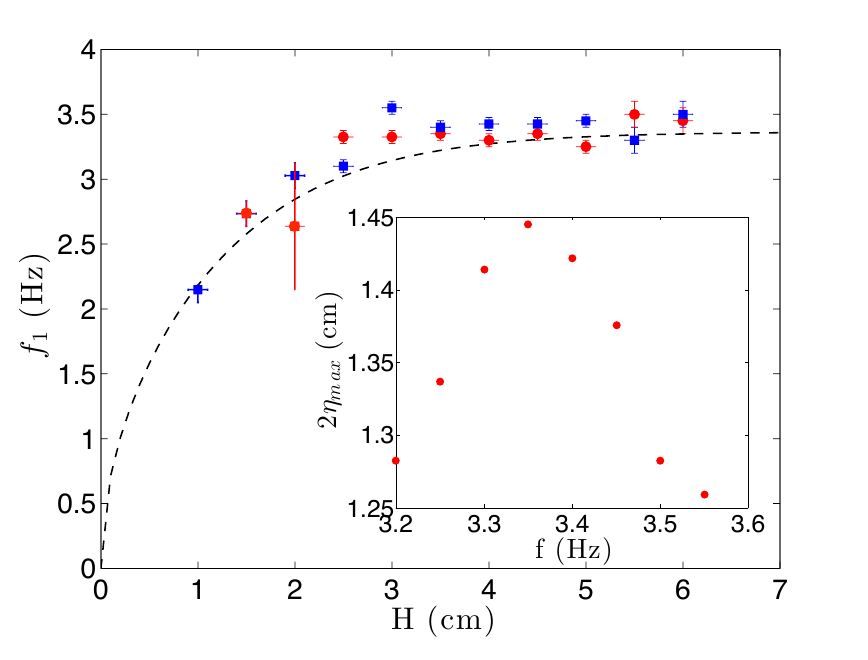}
        \caption{Evolution of the resonant frequency $f_1$ as a function of the water depth $H$ for a container of length $L=7\,{\rm cm}$ and two widths: $w=1.59$ cm (in blue squares) and $w=0.30$ cm (red circles). The dashed line shows the analytical expression of the first resonant frequency $f_1$. Inset: evolution of the maximum amplitude of sloshing $\eta_{max}$ for forcing close to the resonant frequency.}
        \label{fig:appendix_h}
    \end{center}
\end{figure}

Finally, we note that the height of the liquid $H$ is a parameter in the dispersion relation (\ref{appendix_relationdispersion}). The influence of the height of the liquid on the first resonant frequency $f_1$ is reported in figure \ref{fig:appendix_h}. The resonant frequency is obtained from harmonic forcing experiments by estimating the frequency at which the amplitude of sloshing is maximal (inset of figure \ref{fig:appendix_h}). We observe a good agreement between our measurements and the analytical expression. The dispersion relation reaches a plateau for $H \geq 4\,{\rm cm}$. In this situation, the deep water model can be used and the dispersion relation simplifies to ${\omega_n}^{\! 2} = \left(g\,k_n+{\gamma}\,{k_n}^{\! 3}/{\rho} \right)$. Our measurements of the damping coefficient are focused on the deep water regime.

\bibliography{Biblio_Sloshing}

\end{document}